\documentclass[11pt]{article}
\usepackage{amsmath,amsfonts,setspace,cancel,url,hyperref,verbatim,slashed,amsthm,graphicx,color,cite}
\usepackage[a4paper]{geometry}
\newcommand{\be}{\begin{equation}}
\newcommand{\ee}{\end{equation}}
\usepackage[utf8]{inputenc}
\newcommand{\gM}{\mathcal{M}}

\newcommand{\Pd}{P_{\text{Dir}}}
\newcommand{\Pn}{P_{\text{Neu}}}
\newcommand{\dalpha}{\alpha}
\newcommand{\dbeta}{\beta}

\newcommand{\Gthree}{\mathrm{SL}(3) \times \mathrm{SL}(2)}
\newcommand{\Gfour}{\mathrm{SL}(5)}
\newcommand{\Gfive}{{SO}(5,5)}
\newcommand{\Gsix}{{E}_{6(6)}}
\newcommand{\Gseven}{{E}_{7(7)}}
\newcommand{\Geight}{{E}_{8(8)}}
\newcommand{\Aa}{\mathcal{A}}
\newcommand{\Ab}{\mathcal{B}}
\newcommand{\gV}{V}
\newcommand{\cH}{\mathcal{H}}

\newcommand{\ii}{{\bf i}}
\newcommand{\jj}{{\bf j}}
\newcommand{\kk}{{\bf k}}
\newcommand{\lll}{{\bf l}}

\newcommand{\hmu}{\hat{\mu}}
\newcommand{\hnu}{\hat{\nu}}

\begin{document}

\title{
\bf
Open exceptional strings and D-branes
}

\author{\sc Chris D. A. Blair\footnote{\tt cblair@vub.ac.be}}

\date{
Theoretische Natuurkunde, Vrije Universiteit Brussel, and the International Solvay Institutes,  Pleinlaan 2, B-1050 Brussels, Belgium 
}

\maketitle

\begin{abstract}
We study D-branes in the extended geometry appearing in exceptional field theory (or exceptional generalised geometry). 
Starting from the exceptional sigma model (an $E_{d(d)}$ covariant worldsheet action with extra target space coordinates), we define open string boundary conditions. 
We write down Neumann and Dirichlet projectors compatible with the preservation of half-maximal supersymmetry by the brane (building on previous work on the definition of generalised orientifold quotients in exceptional field theory). This leads to a definition of D-branes, plus their S-duals, as particular subspaces of the exceptional geometry, and provides an opportunity to study D-branes in U-fold backgrounds.
\end{abstract}

\tableofcontents

\section{Introduction}

T-duality relates D$p$-branes to D$(p\pm 1)$ branes, interchanging Neumann and Dirichlet boundary conditions on the string worldsheet.
If one uses the doubled approach to the string worldsheet \cite{Duff:1989tf,Tseytlin:1990nb, Tseytlin:1990va,Hull:2004in, Hull:2006va}, an elegant picture emerges whereby all D$p$-branes can be viewed as a single $D$-dimensional brane in the $2D$-dimensional doubled target space: this can then intersect with the $D$-dimensional physical subspace in different number of directions in order to reproduce all standard $p$-branes \cite{Hull:2004in,Lawrence:2006ma,Albertsson:2008gq}. In generalised geometry, \cite{Hitchin:2004ut,Gualtieri:2003dx}, which underlies reformulations of supergravity such as \cite{Coimbra:2011nw} and the related formalism of double field theory (DFT) where the spacetime coordinates are doubled \cite{Siegel:1993xq, Siegel:1993th, Hull:2009mi}, this translates into the statement that D-branes are maximally isotropic subspaces of the doubled tangent bundle \cite{Gualtieri:2003dx,Asakawa:2012px}.
The purpose of this paper is to study the corresponding notion of D-branes in the exceptional geometry that appears in exceptional generalised geometry \cite{Hull:2007zu,Coimbra:2011ky,Coimbra:2012af} and exceptional field theory (ExFT) \cite{Berman:2010is, Hohm:2013pua,Hohm:2013vpa, Hohm:2013uia, Hohm:2014fxa, Hohm:2015xna, Abzalov:2015ega, Musaev:2015ces, Berman:2015rcc}.
We will combine insights from the \emph{exceptional sigma model} \cite{Arvanitakis:2017hwb, Arvanitakis:2018hfn} and from the realisation of \emph{orientifold quotients} in exceptional field theory \cite{Blair:2018lbh}. As O-planes and D-branes appear in type II theories alongside each other, we will use the realisation of the former as fixed points under reflections by $\mathbb{Z}_2 \subset E_{d(d)}$ to write down projectors onto Dirichlet and Neumann directions in exceptional geometry. To ensure that we are describing D-branes, we will require compatibility with a \emph{string charge} or \emph{string structure} that appears in the exceptional sigma model. The crucial underlying feature common to both the orientifold and D-brane projections is compatibility with an $E_{d(d)}$ \emph{half-maximal structure} \cite{Malek:2017njj}: thus we can think of D-branes as defining what we might call half-maximal subspaces of the exceptional geometry.

\subsection{Extended sigma models}

We will begin with a string worldsheet action which corresponds to the doubled sigma model of \cite{Hull:2004in, Hull:2006va} (see also \cite{Lee:2013hma}) and also to the exceptional sigma model of \cite{Arvanitakis:2017hwb,Arvanitakis:2018hfn}. Some notation: $\sigma^A = (\sigma^0,\sigma^1)$ are worldsheet coordinates.
Target space coordinates, which are worldsheet scalars, come in two varieties: ``external'' $X^\mu$, $\mu=1,\dots,n$, and ``extended'' $Y^M$, with the latter sitting in a representation, denoted by $R_1$, of either $O(D,D)$ or $E_{d(d)}$. 
Alongside these, we have also an auxiliary worldsheet one-form $\gV_A^M$. 
The worldsheet inverse metric is $\gamma^{AB}$ and the worldsheet alternating symbol is $\epsilon^{AB}$ with $\epsilon^{01} =-1$. Then we write the action as
\be
\begin{split}
S = - \frac{1}{2} \int d^2\sigma & \,T(\mathcal{M},q) \sqrt{-\gamma} \gamma^{A B}\left( \partial_A X^\mu \partial_B X^\nu g_{\mu\nu} + \frac{1}{2} \gM_{MN} D_A Y^M D_B Y^N\right)
\\ & 
+ \epsilon^{A B} q_{MN} \left( 
\Ab_{\mu\nu}{}^{MN} \partial_A X^\mu \partial_B X^\nu 
	+ \Aa_\mu{}^M D_B Y^N \partial_A X^\mu 
+\partial_A Y^M \gV_B^N \right)
\,,
\end{split} 
\label{esmaction}
\ee
coupling to a background external metric, $g_{\mu\nu}(X,Y)$, a ``generalised'' metric, $\gM_{MN}(X,Y)$, and to generalised gauge fields $\Aa_\mu{}^M(X,Y)$ and $\Ab_{\mu\nu}{}^{MN}(X,Y)$. 
The two-form $\Ab_{\mu\nu}{}^{MN} = \Ab_{\mu\nu}{}^{NM}$ in fact transforms in a particular $O(D,D)$ or $E_{d(d)}$ representation, $R_2$, which lies in the symmetric tensor product of the generalised coordinate representation with itself, $R_2 \subset (R_1 \otimes R_1)_{sym}$. 
We have the string charge $q_{MN} \in \bar R_2$ which appears contracting the multiplet of two-forms in the Wess-Zumino term in \eqref{esmaction}. Finally,we have written $D_A Y^M = \partial_A Y^M + \Aa_\mu{}^M \partial_A X^\mu + \gV_A^M$.

These background fields can depend in principle on any of the extended coordinates $Y^M$ subject to a choice of solution of the section condition, which requires a limited coordinate dependence. This condition can be written as
\be
\partial \otimes \partial |_{\bar R_2} = 0 \,,
\label{sc}
\ee 
i.e. any combination of two derivatives acting on fields or products of fields must vanish when projected into the $\bar R_2$ representation.
It is common to introduce an invariant tensor $Y^{MN}{}_{KL}$ proportional (for low enough $d$) to the projector onto the $R_2$ representation, such that the section condition is often written as $Y^{MN}{}_{PQ} \partial_M \otimes \partial_N = 0$. This so-called Y-tensor appears in the definition of the generalised Lie derivative, $\mathcal{L}_\Lambda V^M = \Lambda^N \partial_N V^M - V^N \partial_N \Lambda^M + Y^{MN}{}_{PQ} \partial_N \Lambda^P V^Q$ \cite{Berman:2012vc}, which defines the local symmetries of the background spacetime, namely $E_{d(d)}$ or $O(D,D)$ valued diffeomorphisms associated to the coordinates $Y^M$ (rather than conventional $\mathrm{GL}(\mathrm{dim} \,R_1)$ diffeomorphisms). A \emph{solution} of the section condition is a choice of physical coordinates $Y^i \subset Y^M$ on which the fields can depend such that \eqref{sc} holds. In exceptional geometries, there is a $d$-dimensional solution (corresponding to 11-dimensional M-theory), and inequivalent $(d-1)$-dimensional solutions corresponding to the 10-dimensional IIA and IIB theories \cite{Hohm:2013pua,Blair:2013gqa}, while in doubled geometry the solutions are $D$-dimensional and again correspond to IIA or IIB.

The string charge $q \in \bar{R}_2$ appearing in the Wess-Zumino coupling of the action \eqref{esmaction} is required in order to write down a coupling to the multiplet of two-forms $\Ab_{\mu\nu} \in R_2$. 
This charge obeys a constraint
\be
q \otimes \partial |_{R_3} = 0 \Leftrightarrow
q_{P K} Y^{KL}{}_{MN} \partial_L  = q_{MN} \partial_P
\label{magic_intro}
\ee 
which should be thought as being solved after solving the section condition for $\partial_M$, and which guarantees gauge invariance of the action.
This charge also appears in the ``tension'', which is given by 
\be
T(\gM,q) \equiv \sqrt{ \gM^{MP} \gM^{NQ} q_{MN} q_{PQ} / 2D}\,,
\label{tension}
\ee
with $D = d-1$ for the $E_{d(d)}$ string. We will henceforth abbreviate $T \equiv T(\gM,q)$.

The final ingredient in \eqref{esmaction} is the auxiliary worldsheet one-form $\gV_A^M$, which is constrained such that $\gV_A^M \partial_M = 0$, again to be thought as being imposed after first solving the section condition.
Integrating out the surviving components of $\gV_A^M$ after solving this constraint eliminates the dual coordinates from the action, imposing a twisted duality constraint relating them to $d-1$ physical coordinates, and reducing the action to the usual action for a string or 1-brane.

It is important to emphasise that the whole action, including the appearance of the auxiliary worldsheet one-form $\gV_A^M$ and the charge constraint \eqref{magic_intro}, follows from gauge invariance, assuming the natural coupling to the two-form $\Ab_{\mu\nu}$ via $q$.
For instance, invariance under the gauge transformation $\delta B_{\mu\nu} = \partial \otimes \Theta_{\mu\nu} |_{R_2}$, where $\Theta_{\mu\nu} \in R_3$, inevitably requires \eqref{magic_intro}.

Let us specify the precise details needed to specify the action \eqref{esmaction} in the more familiar doubled case, and as the exceptional sigma model.
 
\begin{itemize}
\item \emph{Doubled string.} We have $Y^M$ in the vector representation of $O(D,D)$, so that $R_1 = \mathbf{2D}$. The section condition involves a projection onto $R_2 = \mathbf{1}$, and is equivalent to $\eta^{MN} \partial_M \otimes \partial_N = 0$, with $\eta^{MN}$ the inverse of the $O(D,D)$ structure,
\be
\eta_{MN} =\begin{pmatrix} 0 & I \\ I & 0 \end{pmatrix} \,.
\ee
Writing $Y^M = ( Y^i , \tilde Y_i )$, the standard solution to the section condition is that $\partial_M = ( \partial_i, 0 )$, i.e. $\tilde \partial^i = 0$. 
We have $Y^{MN}{}_{PQ} = \eta^{MN} \eta_{PQ}$, and the charge can always be written as $q_{MN} = T_{F1} \eta_{MN}$.
As the generalised metric obeys $\gM_{MN} \eta^{NP} \gM_{PQ} = \eta_{MQ}$, the tension \eqref{tension} reduces to $T = T_{F1}$. 

\item \emph{Exceptional string: $\Gfour$.} This is the case when $n=7$ and $d=4$. The extended coordinates are in the antisymmetric representation of $\mathrm{SL}(5)$, thus we write them as $Y^{ab} = - Y^{ba}$, with $a,b=1,\dots,5$.
We have $R_2 = \mathbf{\bar{5}}$, and the section condition is $\epsilon^{abcde} \partial_{bc} \otimes \partial_{de} = 0$.
The Y-tensor is $Y^{ab,cd}{}_{ef,gh} = 4! \delta^{[abcd]}_{\,\,efgh}$.
The string charge is $q^a$ and obeys
\be
q^b \partial_{ab} = 0 \,.
\ee
One three-dimensional solution of the section condition involves breaking $\mathrm{SL}(5)$ to $\mathrm{GL}(3)$. Letting $a=(i,4,5)$ with $i=1,2,3$, we take $\partial_{ab} = ( \partial_{i5}, 0 )$. 
Then the only allowed charge is $q^i=q^5=0$, $q^4 = T_{F1}$.
This describes a type IIA string with target space coordinates $(X^\mu , Y^{i5})$, after integrating out the non-zero components of the auxiliary one-form.
The tension is $T = T_{F1}$.

Another solution involves breaking $\mathrm{SL}(5)$ to $\mathrm{GL}(3) \times \mathrm{SL}(2)$. Letting $a=(i,\dalpha)$, with $i=1,2,3$ and $\dalpha=4,5$, we take $\partial_{ab} = ( \partial_{ij}, 0 )$.
Then the only allowed charge is $q^i = 0, q^{\dalpha} \neq 0$. 
This describes a type IIB $(p,q)$ string with target space coordinates $(X^\mu , Y^{ij})$, after integrating out the non-zero components of the auxiliary one-form.
The tension is proportional to $T \sim  \sqrt{ \mathcal{H}_{\dalpha \dbeta} q^{\dalpha} q^{\dbeta}}$, where $\mathcal{H}_{\dalpha \dbeta}$ encodes the background dilaton and RR 0-form as an $\mathrm{SL}(2) / \mathrm{SO}(2)$ coset element.

\item \emph{Exceptional string: general results.} In general, the extended coordinates decompose in terms of IIA and IIB physical and dual coordinates as follows:
\be
Y^M = \begin{cases} ( Y^i, \tilde Y_i, \tilde Y , \tilde Y_{ij} , \tilde Y_{ijkl}, \dots ) & \text{IIA} \\
 ( Y^{i}, \tilde Y_{i}{}^{\dalpha}, \tilde Y_{ijk}, \dots ) & \text{IIB} \end{cases} \,.
 \label{Ygeneral}
\ee
In the IIA case, the dual coordinates written here are conjugate to winding modes of the F1 string and D$p$ branes with $p$ even.
In the IIB case, they are conjugate to winding modes of the F1 string and D$p$ branes with $p$ odd: in fact the F1 and D1 winding coordinates appear together as the $\mathrm{SL}(2)$ doublet $\tilde Y_i{}^{\dalpha}$.
There will also be coordinates conjugate to winding modes of the NS5 brane, Kaluza-Klein monopole, and (for high enough $d$) other ``exotic'' branes, denoted by the ellipsis in \eqref{Ygeneral}.
The non-zero components of the charge, assuming the standard 10-dimensional solutions of the section condition, are always:
\be
q_{MN} = \begin{cases} q_{ i }{}^j = q^j{}_i\sim T_{F1} \delta_i{}^j & \text{IIA} \\  q_{ i }{}^{j}{}_{\dalpha} = q^j{}_{\dalpha, i}\sim q_{\dalpha}\delta_i{}^j & \text{IIB} \end{cases}
\label{qsoln}
\ee
(the S-duality $\mathrm{SL}(2)$ indices $\dalpha,\dbeta$ can be raised and lowered using $\epsilon_{\dalpha \dbeta}$). Hence we always obtain the F1 action in IIA, and the $(p,q)$ string action in IIB.
Note that there are no solutions to the charge constraint \eqref{magic_intro} in the 11-dimensional solutions of the section constraint, as there are no strings in M-theory.
\end{itemize}

\subsection{Boundary conditions} 
\label{boundaryconditions} 

The realisation of doubled D-branes using the doubled sigma model was discussed in Hull's paper \cite{Hull:2004in}, and further studied in \cite{Lawrence:2006ma,Albertsson:2008gq}.
We now follow this approach and apply it to the exceptional sigma model \eqref{esmaction}.
For simplicity, we restrict to backgrounds with $\Aa_\mu{}^M = \Ab_{\mu\nu}{}^{MN} = 0$. We will in fact consider the sigma model in terms of the action
\be
\begin{split}
S  &= - \frac{1}{2} \int d^2\sigma \,T \sqrt{-\gamma} \gamma^{A B}\left( \partial_A X^\mu \partial_B X^\nu g_{\mu\nu} + \frac{1}{2} \gM_{MN} 
\partial_A Y^M 
\partial_B Y^N 
\right)
\,,
\end{split} 
\label{esmaction_simple}
\ee
supplemented by the constraint:
\be
T  \sqrt{-\gamma} \gamma^{A B} \gM_{MN} \partial_B Y^N  = \epsilon^{A B} q_{MN} \partial_B Y^N\,. 
\ee
This formulation is equivalent to that where the constraint is implemented by gauging the shift symmetry in dual directions (a consequence of the section condition), leading to the introduction of $\gV_A^M$ as the gauge field for this symmetry \cite{Hull:2004in, Hull:2006va, Lee:2013hma, Arvanitakis:2018hfn}. (Note that one could view this, when the background metrics are flat, as describing the exceptional sigma model on the background $\mathbb{R}^{1,n-1} \times \mathbb{T}^{\mathrm{dim}\,R_1}$, i.e. on an ``exceptional torus''.)
Varying \eqref{esmaction_simple} gives the following boundary terms:
\be
\begin{split} 
\delta S  & \supset - \int d^2 \sigma \, \partial_A \left( T \sqrt{-\gamma} \gamma^{A B} \delta X^\mu \partial_B X^\nu g_{\mu\nu} 
+ \frac{1}{2}  T \sqrt{-\gamma} \gamma^{A B} \delta Y^M \gM_{MN} \partial_B Y^N \right) \\
\end{split} \,.
\ee
Let us now work in conformal gauge, $\gamma^{00} = - 1 = - \gamma^{11}$, $\gamma^{01} = 0$, $\epsilon^{01} = - 1$.
Our interest is in the boundary conditions for the extended coordinates $Y^M$.
For the time being we will assume Neumann boundary conditions for the $X^\mu$, that is $\partial_1 X^\mu=0$ at the worldsheet boundaries at $\sigma=0, \pi$, and comment on the imposition of Dirichlet boundary conditions in these directions later (in section \ref{extsec}).
So we are studying the boundary condition:
\be
\delta Y^M \gM_{MN} \partial_1 Y^N \Big|_{bry} = 0 \,.
\ee
At $\sigma^1 = 0$, let $(\Pd)^M{}_N$ denote the projector onto Dirichlet directions, and let $(\Pn)_N{}^M = \delta_N^M - (\Pd^t)_N{}^M$ denote the Neumann projector.
We have $\Pd \delta Y = 0 = \delta Y \Pd^t$. 
This implies that we have to require
\be
(\Pn)_M{}^N \gM_{NP} \partial_1 Y^P = 0 
\ee
at $\sigma^1 = 0$. (So note the Neumann projector naturally acts on $\gM_{MN} \partial_1 Y^N$, hence its index structure).
Compatibility with the constraint
\be
q_{MN} \partial_0 Y^N =  T\gM_{MN} \partial_1 Y^N \,,
\ee 
then means
\be
( \Pn)_M{}^N q_{NP} \partial_0 Y^P = 0 \,,
\ee
at $\sigma^1 = 0$. This can be achieved if
\be
( \Pn)_M{}^N q_{NP} = q_{MN} ( \Pd)^N{}_P \,,
\label{PqqP}
\ee
which in turn implies
\be
\Pd^t q \Pd = 0 = \Pn q \Pn^t \,.
\label{Pnull}
\ee
Evidently, at the other endpoint, $\sigma^1 = \pi$, we introduce similarly projectors $\widetilde\Pd$ and $\widetilde\Pn$, which need not coincide with the ones at $\sigma^1 = 0$.
Thus each endpoint of the string can be attached to a different subspace of the full extended space. (However, for the rest of this paper, we will assume that both endpoints of the string obey identical boundary conditions.)

When we are dealing with the doubled string, the situation is geometrically appealing.
Note for $O(D,D)$, we have $q_{MN} = \eta_{MN}$ (setting the tension $T_{F1}$ to 1), which is invertible, so that
\be
\Pd = \eta^{-1} \Pn \eta 
\ee
so Dirichlet and Neumann projectors are mapped into each other by applying $\eta$. 
Equivalently, for every Dirichlet direction we have a Neumann direction, reflecting the fact that T-duality interchanges these boundary conditions.
A doubled D-brane then amounts to a $D$-dimensional subspace of the $2D$-dimensional doubled space, and the canonical form of the projectors $\Pd$ and $\Pn$ is
\be
\Pd = \begin{pmatrix} 0 & 0 \\ 0 & I \end{pmatrix} 
\,,\quad
\Pn = \begin{pmatrix} I & 0 \\ 0 & 0 \end{pmatrix} \,.
\ee
Depending on how one chooses the physical coordinates, the doubled D-brane will intersect with the $D$-dimensional \emph{physical} subspace in differing numbers of directions, and so realises the full set of expected $p$-branes.

For the exceptional sigma model, the string charge $q_{MN}$ will not be invertible. 
The ``pairing'' between Dirichlet and Neumann directions implied by \eqref{PqqP} is then not fully determined.
The constraint \eqref{magic_intro} on the charge implies that it always takes the form \eqref{qsoln}, so that we basically have 
\be
q_{MN} = \begin{pmatrix} 0 & I & 0 \\
I & 0 & 0 \\
0 & 0 & 0 
\end{pmatrix}
\,(\text{IIA})
\,,\quad  
q_{MN} = \begin{pmatrix} 0 & pI & qI & 0 \\
pI & 0 & 0 & 0 \\
qI & 0 & 0 & 0 \\
0 & 0 & 0 & 0 
\end{pmatrix} 
\,(\text{IIB})
\,.  
\label{qhere} 
\ee
Note that (when either $q=0$ or $p=0$ in the IIB case) this is an embedding of the $SO(D,D)$ structure into $E_{d(d)}$ language (breaking the latter to the former).
In order to find Neumann and Dirichlet projectors obeying \eqref{PqqP} for $q_{MN}$ of the form \eqref{qhere}, we will use some additional information.

\subsection{O, D}

String theory also contains orientifold planes, which are (non-dynamical) extended objects carrying (negative) RR charge, and which appear alongside D-branes of the same dimensionality (as required for charge cancellation). In particular IIA contains O$p$ planes and D$p$ branes with $p$ even, while in IIB we have $p$ odd (we only consider stable $p$-branes).
An elegant description of orientifold quotients (at the supergravity level) in exceptional field theory was developed in \cite{Blair:2018lbh}.
For the standard orientifolds, we consider a quotient by $\mathbb{Z}_2 \subset E_{d(d)}$, with this $\mathbb{Z}_2$ acting ``geometrically'' on the fields and coordinates of exceptional field theory according to how they transform as representations of $E_{d(d)}$.
The fixed points of this $\mathbb{Z}_2$ define generalised orientifold planes as subspaces of the extended geometry. 
These intersect with the physical geometry (defined by the choice of coordinates solving the section condition) in order to realise spacetime orientifold planes of each dimension: in addition, one obtains descriptions of the heterotic supergravities and of orbifolds of M-theory (including the Ho\v{r}ava-Witten description of M-theory, or 11-dimensional supergravity, on an interval).
It is clear that in the section condition solutions in which the generalised orientifold describes genuine type II orientifolds, that the fixed point planes coincide with the subgeometry that should be spanned by an exceptional notion of a D-brane.

The connection can be formalised using supersymmetry. Exceptional field theory describes maximal supergravity in 11 dimensions and lower (see for example \cite{Godazgar:2014nqa}).
In order to describe backgrounds which break some supersymmetry, or truncations to theories with less supersymmetry, an $E_{d(d)}$ covariant notion of a \emph{half-maximal structure} can be defined \cite{Malek:2017njj}.
This is a set of generalised tensors, globally defined on the physical spacetime underlying the exceptional field theory construction, obeying certain compatibility conditions, whose existence is equivalent to that of a set of Killing spinors implying the presence of half-maximal supersymmetry. 

The generalised orientifold quotients (or ``O-folds'') considered in \cite{Blair:2018lbh} are restricted by the requirement that they preserve the existence of the $E_{d(d)}$ half-maximal structure, and thus lead to configurations with half the supersymmetry. 
The important point for us now is that D-branes themselves are of course half-BPS objects; this underlies how they can appear alongside O-planes in half-maximal theories (type I and its T-duals). 

Putting O and D together, we propose that we can use the $\mathbb{Z}_2$ transformation of \cite{Blair:2018lbh} to define the correct Dirichlet and Neumann projectors obeying \eqref{PqqP}, and which describe therefore D-branes as ``half-maximal subspaces'' in the exceptional geometry of ExFT.

\section{A definition of D-branes in exceptional geometry}

\subsection{D-brane structure}
\label{defD}

Now we give a formal definition of what we might choose to call a \emph{D-brane structure} in exceptional geometry.
By exceptional geometry we mean either that appearing in exceptional field theory \cite{Berman:2010is, Hohm:2013pua} or alternatively in exceptional generalised geometry \cite{Hull:2007zu,Coimbra:2011ky,Coimbra:2012af}.
In exceptional field theory, we have fields depending on the extended coordinates $(X^\mu, Y^M)$, and generalised vectors and tensors transforming in the representations, $R_1, R_2, R_3, \dots$ of $E_{d(d)}$. 
In exceptional generalised geometry, we work with a generalised tangent bundle $E$ over a base manifold $M$, and this generalised tangent bundle carries an action of $E_{d(d)}$.
To be precise, here we would take $M$ to be a $(d-1)$-dimensional manifold and define a set of bundles $\mathcal{R}_1$, $\mathcal{R}_2$, $\mathcal{R}_3,\dots$, such that generalised tensors transforming in the $R_p$ representations of $E_{d(d)}$ are sections of these bundles. For instance, the generalised tangent bundle itself is $E = \mathcal{R}_1$ with
\be
\mathcal{R}_1 \simeq  TM \oplus T^*M \oplus \Lambda^5T^*M \oplus ( T^*M \otimes \Lambda^6 T^*M) \oplus\Lambda^{\text{even/odd}} T^* M
\ee
with the even/odd antisymmetric products corresponding to IIA and IIB respectively, while
\be
\mathcal{R}_2 \simeq \mathbb{R} \oplus \Lambda^4 T^*M \oplus (T^*M \otimes \Lambda^5 T^*M ) \oplus \Lambda^{\text{odd/even}} T^* M \oplus \dots 
\ee
where the ellipsis denotes additional factors needed for $d=7$ \cite{Grana:2009im}.
We mentioned earlier that the representation $R_2$ is contained within the symmetric tensor product of $R_1$ with itself.
This fact allows us to define a (symmetric) product $\wedge : \mathcal{R}_1 \times \mathcal{R}_1 \rightarrow \mathcal{R}_2$ 
which takes a pair of sections of $\mathcal{R}_1$ and projects them into a section of $\mathcal{R}_2$, which we will use below.

We can think of the extended geometry of ExFT as being locally isomorphic to the extended tangent bundle $E$.
We will therefore describe our D-brane structure in terms of maps on $E$.
In both cases, we will write partial derivatives $\partial_M$. In ExFT, we think of the choice of solution of the section condition as telling us which components of these are non-zero, corresponding to derivatives with respect to the physical coordinates. Then different choices of this solution correspond to IIA versus IIB. In exceptional generalised geometry, we think of the physical coordinates of the underlying manifold $M$ as being embedded into $\partial_M$ with all other components zero. Then different choices of this embedding are used for IIA versus IIB.

The data we use to specify a D-brane structure in exceptional geometry consists of:
\begin{itemize} 
\item an involution $Z : E \rightarrow E$, $Z^2 = 1$, which defines projectors
\be
\Pn^t = \frac{1}{2} ( 1 + Z ) \,,\quad 
\Pd = \frac{1}{2} ( 1 - Z) 
\ee
\item a section of the $\mathcal{\bar{R}}_2$ bundle, $q$, obeying the \emph{string charge condition} 
\be
q \otimes \partial |_{{R}_3}  = 0 \,.
\label{magicrep}
\ee
We can use this to define a degenerate bilinear form
\be
q : E \otimes E \rightarrow \mathbb{R} \,,\quad
(U,V) \mapsto q(U,V) \,,
\ee
(which we could also see as a non-invertible map from the exceptional tangent bundle $E$ to its dual $E^*$, $q : E \rightarrow E^*$) using the symmetric map $\wedge: \mathcal{R}_1 \otimes \mathcal{R}_1 \rightarrow \mathcal{R}_2$, so that $q(U,V) \equiv q \cdot ( U \wedge V)$ where on the right hand side we use the natural pairing between sections of $\mathcal{R}_2$ and $\mathcal{\bar R}_2$, denoting this by a dot.
We require that (this is the condition \eqref{PqqP} arising from the worldsheet boundary conditions and self-duality constraint)
\be
q( \Pn^t U, V ) = q ( U, \Pd V ) \,,
\ee
for arbitrary $U,V \in \Gamma(E)$, 
or equivalently that
\be
q(Z U, Z V ) = - q ( U, V ) \,.
\label{qZZ}
\ee
This implies that (this is \eqref{Pnull})
\be
q(\Pn^t U , \Pn^t V) = 0 = q(\Pd U , \Pd V) \,,
\ee
i.e. both the images of the projectors are null with respect to the string charge $q$.

\item an $E_{d(d)}$ half-maximal structure \cite{Malek:2017njj}, consisting of $d-1$ generalised vectors $J_u \in \Gamma(E)$, $u=1,\dots, d-1$, and generalised tensors, $K \in \Gamma(\mathcal{R}_2)$, $\hat K \in \Gamma(\mathcal{\bar{R}}_2)$, obeying certain compatibility conditions, and such that they are contained in the image of the Neumann projector,
\be
\Pn^t J_u = J_u \,\quad
\Pn  K = K\,,\quad
\Pn  \hat K = \hat K \,,
\label{preserveJ}
\ee
i.e. they are invariant under the involution.
Note the compatibility conditions include $K \cdot \hat K > 0$ and $J_u \wedge J_v = \delta_{uv} K$. 
As we have firstly that $q(J_u, J_v) = 0$, it also follows that $q \cdot ( J_u \wedge J_v ) = \delta_{uv} q \cdot K = 0$. Roughly speaking, both $q$ and $(K, \hat K)$ define separate $SO(D,D)$ structures whose intersection determines the orientation of the D-brane in the physical subspace. We have some further comments on this in appendix \ref{extra}.

\end{itemize}

As in \cite{Lawrence:2006ma,Albertsson:2008gq}, we can also require the Dirichlet and Neumann projectors to be orthogonal with respect to the generalised metric, and that the Neumann subbundle is integrable, i.e. that $\Pd \mathcal{L}_{\Pn^t U} \Pn^t V = 0$ for all $U,V$, where $\mathcal{L}$ is the generalised Lie derivative. 

In components, given the transformation $Z^M{}_N$ squaring to the identity, the crucial conditions \eqref{magicrep} and \eqref{qZZ}, become  
\be
q_{M K} Y^{K L}{}_{P Q} \partial_L = q_{P Q} \partial_M \,,
\label{magic}
\ee
\be
Z^M{}_P Z^N{}_Q q_{MN} = - q_{PQ} 
\,,
\label{Zq}
\ee
and the preservation of the half-maximal structure $J_u{}^M$, $K^{MN}$, $\hat K_{MN}$ is that
\be
Z^M{}_N J_u{}^N = J_u{}^M \,,\quad
Z^P{}_M Z^Q{}_N  K^{MN} =  K^{PQ} \,,\quad
Z^M{}_P Z^N{}_Q \hat K_{MN} = \hat K_{PQ} \,.
\label{Zhalfmax}
\ee
In practice, we can find $Z^M{}_N$ as in \cite{Blair:2018lbh} by picking a suitable form for the half-maximal structure and working out the action of its stabiliser subgroup within $E_{d(d)}$.
Then we specialise to a $\mathbb{Z}_2$ discrete subgroup of this stabiliser. 
This was worked out explicitly for $\Gfour$ and $\Gfive$ in \cite{Blair:2018lbh} but applies for higher rank groups too (note that the details of the half-maximal structure are slightly different in $d=7$ \cite{Malek:2017njj}). 
In the next subsection, we will use the results on $\Gfour$ to explore how the above definitions work out in an explicit example.

First, we can already give some general expressions. 
In particular, we can write down $Z^M{}_N$ explicitly acting on $E$ in its decomposition into $O(D,D)$ (really $SO(D,D)$) representations by using the results of appendix B of \cite{Blair:2018lbh}. 
Decomposing $E_{d(d)}$ to $SO(D,D)$ with $D=d-1$, we have $R_1 = \mathbf{2D} \oplus \mathbf{2^{D -1}} \oplus \mathbf{r}$, where for $d<6$ the final representation $\mathbf{r}$ is not present, for $d=6$ it is the trivial representation, and for $d=7$ it is another copy of the fundamental. The representation $\mathbf{2^{D-1}}$ is the Majorana-Weyl spinor representation of (the double cover of) $SO(D,D)$.
This spinor can be viewed as the formal sum of even or odd $p$-forms in spacetime, depending on its chirality. The case where it corresponds to even $p$-forms is IIA, and the odd $p$-forms give IIB. (This is the opposite chirality to the RR gauge fields themselves: this is because the $R_1$ representation corresponds to the gauge transformation parameters of these fields.)

An explicit realisation of this (as in \cite{Hohm:2011dv}) involves introducing creation and annihilation operators $(\psi^i, \psi_i)$ (with $i=1,\dots,D$) obeying $\{ \psi^i, \psi_i \} = \delta^i_j$, $\{\psi^i, \psi^j \} = 0$, $\{ \psi_i, \psi_j \} = 0$. 
Defining a vacuum $|0\rangle$ such that $\psi_i |0\rangle = 0$, we build spinors of definite chirality by acting with an even or odd number of $\psi^i$ on $|0\rangle$. 
Then when the $\mathbb{Z}_2$ transformations acts on the doubled vector representation $\mathbf{2D}$ as $Z = \text{diag} ( I_p, -I_n, -I_p, I_n)$ (corresponding to $n$ Dirichlet directions and $p$ Neumann directions in the $D$-dimensional physical space), it acts on the spinor as the operator \footnote{This is related to the operator $\tilde Z$ in \cite{Blair:2018lbh} by $\hat Z = i \tilde Z$. This is because $\tilde Z$ acted on the spinor corresponding to the RR fields themselves, while $\hat Z$ acts on the spinor $\mathcal{C}$ corresponding to the extended coordinates or equivalently to the gauge transformation parameters $\lambda$ of the RR fields, which have opposite chirality. The gauge transformation is $\delta \mathcal{C} = \sqrt{2} \psi^i \partial_i \lambda$, and we have from \cite{Blair:2018lbh} that $\tilde Z \psi^i = i \psi^i \tilde Z$. }
\be
\hat Z \equiv (-1)^{\frac{1}{2} ( \tilde N - 1) } \,,\quad
\tilde N \equiv \sum_{\mu=1}^p\psi^\mu \psi_\mu 
+ \sum_{a=1}^n \psi_a \psi^a 
=N_{(p)} - N_{(n)} + n \,,
\ee
where we split $i=(\mu, a)$ corresponding to the even and odd physical directions, and $N_{(p)}$ and $N_{(n)}$ denote the number operators for the $(\psi_\mu, \psi^\mu)$ and $(\psi_a, \psi^a)$ spinor subspaces. On a spinor state
\be
\chi\equiv \frac{1}{m! q!}C_{\mu_1 \dots\mu_m a_1 \dots a_q} 
\psi^{\mu_1} \dots \psi^{\mu_m} \psi^{a_1} \dots \psi^{a_q} |0\rangle 
\ee
we have 
\be
\hat Z \chi = (-1)^{(m+n-q-1)/2} \chi \,.
\ee
Note that the action of $Z$ and $\hat Z$ does not correspond to a transformation in $O(D,D)$ or its double cover. In particular, $Z$ sends the $O(D,D)$ structure $\eta_{MN}$ to $-\eta_{MN}$. Despite this, they represent a symmetry of the doubled geometry, preserving the action and local symmetry transformations (in which $\eta_{MN}$ appears alongside its inverse: the combination of the pair is invariant under the action of $Z$). Though $Z$ always squares to one, we have $\hat Z^2 = (-1)^{\tilde N - 1}$.
As $\tilde N = N_F + n - 2 N_{(n)}$, $\hat Z^2 =1$ only if $N_F + n -1$ is even, hence if $N_F$ is even/odd then $n$ is odd/even. This picks out the conventional spinor and $p$-form chiralities along with the correct D-brane dimensions in IIA and IIB.

By taking $n=0$, so that $Z = \text{diag}(I_D, -I_D)$ on the $\mathbf{2D}$ representation, we obtain $\hat Z = (-1)^{\frac{1}{2}(N_F -1)}$, where $N_F$ is the total number operator. This must be odd, so this is an action in IIB (in particular defining 9-branes). Then it is easy to see that $\tilde Z$ acts as $+1$ on the $p$-forms with $p=1,5,9$ and as $-1$ on the $p$ forms with $p=3,7$. It follows that acting on $\mathbf{2D} \oplus \mathbf{2^{D-1}}$ there are thus $D + {D \choose 3} + {D \choose 7}$ Dirichlet directions.

\begin{table}[ht]
\centering
	\begin{tabular}{lc|ccc|c}
		$E_{d(d)}$ &$R_1$  & $\mathbf{2D}$ & $\mathbf{2^{D-1}}$ & $\mathbf{r}$  &Total \\\hline 
		$\Gthree$ & $(\bf 3,2)$ & 2 & 0 &- & 2\\
		$\Gfour$ & $\bf 10$ & 3 & 1 & - & 4\\
		$\Gfive$ & $\bf 16$ & 4 & 4 & -& 8\\
		$\Gsix$ & $\bf 27$ & 5 & 10 & 1 & 16\\
		$\Gseven$ & $\bf 56$ & 6 & 20 & 6 & 32\\
	\end{tabular}
	\caption{Number of Dirichlet directions in the decomposition of $R_1$ into $O(D,D)$ representations}
	\label{nos}
\end{table}

The action of the $\mathbb{Z}_2$ transformation on the remaining representation ${\bf r}$ which appears for $d=6$ and higher can be found starting with \cite{Blair:2018lbh}. A direct if unimaginative route, therefore taken naturally by the present author, uses the information there about the $\mathbb{Z}_2$ appearing in the Ho\v{r}ava-Witten description of M-theory on an interval. As we know how this acts on all the fields in the $\mathrm{SL}(5)$ ExFT, we learn how it acts in spacetime on the metric, three-form, dual six-form and also on the dual graviton (which appears in certain $E_{d(d)}$ representations).\footnote{Specifically, $g$ is even, $C_3$ is odd, $C_6$ is even and $h_{8,1}$ is odd.} Then we can reverse engineer the $\mathbb{Z}_2$ transformation in the groups $E_{d(d)}$ for $d>4$ and count the number of minus signs. This is conveniently done acting on the field $\mathcal{A}_\mu{}^M$ at the start of the ExFT tensor hierarchy, which is $R_1$ valued. This reveals that for $\Gsix$, the $\mathbb{Z}_2$ must end up acting as $-1$ on the representation $\mathbf{r} = \mathbf{1}$, and for $\Gseven$ it must act as it does on the $\mathbf{2D}$ representation on $\mathbf{r} = \mathbf{2D}$. 
For $\Geight$, a large number of additional dual fields make an appearance, so this method does not immediately tell us the answer. We can note however the general pattern is that there are always $2^{D-1}$ Dirichlet directions, as shown in table \ref{nos}. 

Let's take stock of the general situation before we move on to an explicit example in $\Gfour$. 

We have a $\mathbb{Z}_2$ transformation acting on the exceptional geometry of ExFT or generalised geometry, corresponding to that used originally in \cite{Blair:2018lbh} to define a generalised orientifold quotient. This $\mathbb{Z}_2$ preserves half-maximal supersymmetry. We can use it to define a pair of projectors, $\Pd = \frac{1}{2} ( 1 - Z)$ and $\Pn^t = \frac{1}{2} ( 1 + Z)$, which can be used to define Dirichlet and Neumann boundary conditions for the exceptional sigma model. Consistency of these boundary conditions brings into play the string charge $q_{MN}$, and we must have $Z^P{}_M Z^Q{}_N q_{PQ} = - q_{MN}$. Allowed pairs $(q,Z)$ obeying this compatibility condition pick out D-branes as ``half-maximal subspaces'' of the exceptional geometry. 

If we view the target space of the exceptional sigma model as an ``exceptional torus'', then the D-branes defined in this way wrap part of this torus.
For $\Gfour$, the branes wrap a $T^6 \subset T^{10}$, for $\Gfive$ a $T^8 \subset T^{16}$, for $\Gsix$ a $T^{11} \subset T^{27}$ and for $\Gseven$ a $T^{24} \subset T^{56}$. In addition, they wrap either the entire $(11-d)$-dimensional external space or a subspace thereof (see the discussion in section \ref{extsec}). Overlapping the space wrapped by the brane with the physical space selected by the choice of section condition allows branes of different dimensionality in spacetime to appear from the same object in the exceptional geometry. 

Finally, we should note that the projectors $\Pd$ and $\Pn$ can be used to define these half-maximal subspaces wrapped by branes also when the compatibility condition with $q$ is not obeyed. In this case, the branes would not be interpreted as D-branes in spacetime. Rather, they may be viewed as some sort of NSNS brane related to the existence of the heterotic theories, as suggested in \cite{Bergshoeff:1998re}, or indeed as the Ho\v{r}ava-Witten end-of-the-world branes \cite{Horava:1995qa, Horava:1996ma}, as implied by the generalised orientifold analysis of \cite{Blair:2018lbh}.

\subsection{Example: $\mathrm{SL}(5)$}

The $d=4$ $\mathrm{SL}(5)$ exceptional geometry \cite{Berman:2010is, Berman:2011cg, Blair:2013gqa, Musaev:2015ces} is an instructive example, and one in which it is simple to enumerate all possibilities. 
We denote five-dimensional fundamental indices by $a,b$.
The $\mathbb{Z}_2$ involution obtained in \cite{Blair:2018lbh} (and which we know from the discussion there is compatible with the existence of the half-maximal structure) can be taken to be:
\be
Z^a{}_b = \text{diag} ( - \delta^{\ii}_{\jj} , + 1 ) \,.
\label{Zab} 
\ee
where $\ii, \jj=1,\dots,4$ label the four odd components of generalised tensors transforming in the $\mathbf{5}$ or $\mathbf{\bar 5}$.
Thus we write $a=(\ii,s)$ with $Z^s{}_s = +1$.
The $R_1$ representation is the antisymmetric, thus generalised vectors $V^M$ are written as $V^{ab}$ with $V^{ab} = - V^{ba}$, and we have $Z^M{}_N$ given by
\be
Z^{ab}{}_{cd} \equiv 2 Z^{[a}{}_c Z^{b]}{}_d \,,
\ee
and projectors
\be
(\Pn^t)^{ab}{}_{cd} = \frac{1}{2} \left(2 \delta^{[a}_c \delta^{b]}_d + 2Z^{[a}_c Z^{b]}_d \right) \,,\quad
(\Pd)^{ab}{}_{cd} = \frac{1}{2} \left(2 \delta^{[a}_c \delta^{b]}_d - 2 Z^{[a}_c Z^{b]}_d \right) \,.
\ee
These can therefore be written in the canonical form 
\be
Z^M{}_N = \begin{pmatrix} 
2\delta^{[\ii}_{\kk} \delta^{\jj]}_{\lll} & 0 \\ 0 & - \delta^{\ii }_{\jj } 
\end{pmatrix} \,,\quad
(\Pn^t)^M{}_N = \begin{pmatrix} 2\delta^{[\ii}_{\kk} \delta^{\jj]}_{\lll} & 0 \\ 0 & 0 \end{pmatrix} \,,\quad
(\Pd)^M{}_N = \begin{pmatrix} 0 & 0 \\ 0 & \delta^{{\bf i} }_{{\bf j}} \end{pmatrix} \,.
\ee
(Note that our contraction convention is $V^M U_M \equiv \frac{1}{2} V^{ab} U_{ab} = \frac{1}{2} V^{\ii \jj} U_{\ii \jj} + V^{\ii s} U_{\jj s}$.)
This means that the components $V^{\ii \jj}$ of a generalised vector $V^M$ are even under $Z$, while the components $V^{\ii s}$ are odd.

The string charge or string structure is $q^a$ or $q_{ab,cd} = \epsilon_{abcde} q^e$. The condition \eqref{Zq} imposes that $q^s =0$ or equivalently $q_{\ii \jj , \kk \lll }=0$.
The condition \eqref{magic} requires that the physical coordinates are embedded into $\partial_{ab}$ such that
\be
q^b \partial_{ab} = 0 \Rightarrow q^{\ii} \partial_{a \ii} = 0 \,.
\ee
We analyse what D-branes are possible by looking in turn at IIA and IIB embeddings, and seeing what are the consequences for writing down $Z^a{}_b$ as in \eqref{Zab} with different choices of the even index $s$ such that $Z^s{}_s = +1$. We can either view this as fixing the involution $Z^a{}_b$ used to define the brane, and changing the choice of section (by taking different decompositions of the $\mathrm{SL}(5)$ index $a$ into $\mathrm{GL}(3)$ indices), or equivalently as fixing the choice of section and taking all possible choices of $Z^a{}_b$. Either point of view is compatible with the idea that we have a single extended notion of brane, whose intersections with the physical subspace realises the full spectrum of standard branes: this is maybe more manifest in the former picture (where we fix the brane definition and rotate the section). However, we will adopt the language of the latter in practice, fixing our notation for the physical solution of the section condition and changing the order of the indices in $Z^a{}_b$.
In terms of the extended geometry, we think of the branes as fixed in the $Y^{\ii s}$ directions and therefore spanning the $(X^\mu, Y^{\bf i \bf j})$ directions.

\begin{itemize}
\item {IIA: let $a=(i,4,5)$ with $s=5$, $\ii = (i,4)$.} In this and all IIA cases, the non-vanishing components of $\partial_M$ are $\partial_{i5}$. 
The string charge $q^i = ( 0 , q^4, 0)$ obeys the defining conditions.
Therefore we obtain a D-brane, with the Dirichlet projector acting on spacetime vectors $V^{i5}$ as $-I$. We may also note that the coordinates $Y^{\ii \jj}$ are all dual coordinates.
Thus the D-brane is extended in the directions $X^\mu$ alone: this is therefore a D6 brane.

\item {IIA: let $a=(i,4,5)$ with $s=4$, $\ii=(i,5)$.} The string charge is forced to vanish: this is not a valid definition of a D-brane in type II theories. Referring to the classification in \cite{Blair:2018lbh}, we see that in fact the $Z$ transformation here would lead to the heterotic $E_8 \times E_8$ theory when applied as a quotient. There are no D-branes in the heterotic theories, so this is consistent.
Another way to say the same thing is to note that the physical directions $(X^\mu, Y^{i5})$ are all even and therefore the brane in this case would be spacetime filling, but IIA does not have D9 branes. One can also see that the M-theory direction $Y^{45}$ is odd under the $\mathbb{Z}_2$ - in the orientifold picture, this is the Ho\v{r}ava-Witten interval.

\item {IIA: let $a=(i,4,5)$ with $s=i$ for one of the $i$, say $i=3$, so $\ii=(1,2,4,5)$.} The string charge $q^i = ( 0 , q^4, 0)$ obeys the defining conditions. Again we obtain a D-brane.
The Dirichlet projector acts on spacetime vectors $V^{i5} = (V^{15}, V^{25}, V^{35})$ as $\text{diag}(+1,+1,-1)$. This means that Dirichlet boundary conditions apply in one direction in spacetime: the branes extend in the $(X^\mu, Y^{15}, Y^{25})$ directions and are therefore D8 branes. 

\item {IIB: let $a=(i,\dalpha)$ with $s$ one of the $\dalpha$ so that $\ii=i$ and the other $\dalpha$ index.} In this and the other IIB case, the non-vanishing components of $\partial_M$ are $\partial_{ij}$.
The string charge is $q^a = ( 0 , q^{\dalpha})$, and so has only one component.
The Dirichlet projector acts on the spacetime vectors trivially. Thus the branes are spacetime filling, extending in the $(X^\mu, Y^{ij})$ directions.
This corresponds to D9 branes, and their S-duals. 
To be precise, when $q^{\dalpha} = (q,0)$, corresponding to $Z^a{}_b = \text{diag}(-1,-1,-1,-1,+1)$ we obtain F1 strings and D9 branes, while when $q^{\dalpha} = (0,q)$, corresponding to $Z^a{}_b = \text{diag}(-1,-1,-1,+1,-1)$ we obtain D1 branes and NS 9-branes.
Interestingly, the S-dual picture suggests that what we describe is a construction of the heterotic string in terms of open D1 branes ending on NS 9-branes, as in \cite{Hull:1998he}.

\item {IIB: let $a=(i,\dalpha)$ with $s$ one of the $i$, say $i=3$, so $\ii = (1,2,\dalpha)$.} The string charge is $q^a = (0, q^{\dalpha})$ and can have both components of $q^\alpha$ non-zero.
The Dirichlet projector acts on spacetime vectors $V^{ij} = (V^{12}, V^{13}, V^{23})$ as $\text{diag}(+1,-1,-1)$. 
So Dirichlet boundary conditions would apply in two directions in spacetime, defining a brane extended in the $(X^\mu, Y^{12})$ directions: therefore a D7.

\end{itemize}

\subsection{Dirichlet conditions in external directions}
\label{extsec} 

We assumed in section \ref{boundaryconditions} that the external directions $X^\mu$ obeyed Neumann boundary conditions, so that the branes we are considering span the entire external space of the full exceptional geometry $(X^\mu, Y^M)$.
Naively, it might seem that one could instead have the $X^\mu$ obey a mix of Neumann and Dirichlet conditions, resulting in $p$-branes for arbitrary $p$ in both the IIA and IIB embeddings!
To rule out the wrong $p$ branes in each case, it is likely that one needs to return to the condition of half-maximal supersymmetry anew, and check what happens when acting with an additional $\mathbb{Z}_2$ reflection on an external direction. We will not attempt this analysis but rather offer one proposal to obtain the correct branes. 

We suppose we pick a single $X^\mu$ direction to be Dirichlet. The interchange of Neumann and Dirichlet boundary conditions could be seen as the result of T-dualising in this direction.
We want to implement this interchange of IIA and IIB directly on the exceptional geometry.\footnote{A fun experiment is also to perform the T-duality on the exceptional sigma model worldsheet action \eqref{esmaction} itself, resulting in some complicated transformation of the background fields whose meaning is not immediately clear (nor immediately helpful in the present circumstances), and in fact may be one step towards enlarging $E_{d(d)}$ to $E_{d+1(d+1)}$.} The first observation towards this end is that the T-duality swapping IIA and IIB is \emph{not} an $E_{d(d)}$ transformation, as $E_{d(d)}$ contains only the $SO(d-1,d-1)$ \emph{symmetry} of IIA and IIB separately as a subgroup. However, after decomposing $Y^M$ into $\mathrm{GL}(d-1)$ representations, one can identify separate sets of IIA and IIB physical coordinates and view the exchange of these coordinates as an \emph{outer automorphism} as in \cite{Malek:2015hma}.
For example, for $\Gfour$, we decompose $Y^{ab} = ( Y^{i5}, Y^{ij} , Y^{i4} , Y^{45})$ and the outer automorphism acts by swapping $Y^{i5} \leftrightarrow Y^{ij}$.
Let us call this transformation $\sigma$.
We propose that switching a single $X^\mu$ from Neumann to Dirichlet corresponds to acting on the exceptional geometry with the transformation $\sigma$, $Y \mapsto \sigma(Y)$.
Thus, in particular, denoting $Y^M = (Y^{(A)}, Y^{(B)}, \tilde Y)$, the diagonal $\mathbb{Z}_2$ transformation $Z = \text{diag}( Z^{(A)}, Z^{(B)}, \tilde Z)$ becomes $Z^\prime \equiv \sigma Z \sigma^{-1} = \text{diag} ( Z^{(B)}, Z^{(A)}, \tilde Z)$.

Then we have 
\be
\Pd^\prime = \frac{1}{2} \begin{pmatrix} 
	1 - Z^{(B)} & 0 & 0 \\ 0 & 1 - Z^{(A)} & 0 \\ 0 & 0 & 1 - \tilde Z
\end{pmatrix} 
\ee
so that in the theory with coordinates $Y^\prime$ the number of Dirichlet directions in the $Y^{\prime (A)}$ IIA physical directions is equal to the number of Dirichlet directions in the original $Y^{(B)}$ IIB physical directions, and vice versa.

We also need to consider the action of $\sigma$ on the string charge $q$. 
To be able to immediately carry across the branes determined by the definition when all external directions are Neumann, we require that $q$ be invariant under $\sigma$.
On the IIB side, this means restricting the IIB charge $q^\alpha$ to correspond to F1 strings alone. 
Then $q_{MN}$ takes the form \eqref{qhere} with $p\neq 0$, $q=0$, and is unchanged on swapping the IIA and IIB coordinates. 
For instance for $\mathrm{SL}(5)$ one has $q_{ij ,k5} = q_{k5, ij} = \epsilon_{ijk}$,\footnote{Here and below we denote by $\epsilon_{ijk}$ the alternating symbol with $\epsilon_{123} =1$.} invariant under $Y^{ij} \leftrightarrow Y^{i5}$.     
By eliminating the S-duals of the original IIB D-branes from the set of possibilities, we ensure that we find only the expected D-branes of each dimension.\footnote{In principle, suppose we start with the D7 and consider its S-dual 7-brane, on which D1 branes end. T-dualising this along worldvolume directions will lead to $p$-branes, $p<7$, which are not D-branes (they will depend on the string coupling $g_s$ as $g_s^{\alpha}$ for $\alpha < -1$), and on which D$p^\prime$ branes, $p^\prime > 1$, end. These will not be found when our string charge $q_{MN}$ obeys the constraint \eqref{magic} and corresponds to 1-branes in 10-dimensions. However, on assuming isometries we may be able to describe more general branes ending on branes - see the comment in the final discussion - which should follow from relaxing the restriction here to $q$ invariant under $\sigma$.}

Clearly if we then impose Dirichlet boundary conditions on a second external coordinate, we apply $\sigma$ again, but $\sigma^2 = 1$, so this brings us back to the original situation. Hence the results of the previous subsection for D$p$ branes (but \emph{not} the S-duals on the IIB side) in $\Gfour$ can be interpreted as holding ``modulo 2''.
In this way, we can indeed define D-branes with any number of external Dirichlet directions: when this number is odd, we have to use the additional symmetry $\sigma$.

\section{D-branes in S- and U-folds} 

We will now apply our definition of D-branes in the $\Gfour$ ExFT to the situation where we have some non-trivial U-duality monodromy, and want to know which D-branes are compatible with this monodromy.
This is a step towards understanding D-branes in U-folds, and generalises the T-fold analysis of \cite{Lawrence:2006ma,Albertsson:2008gq}.
As a proof of concept, we will focus on some illustrative examples in the IIB case, and leave an exhaustive classification for future work.

\subsection{$\mathrm{SL}(5)$ duality in IIB} 

We will focus on the 10-dimensional extended geometry of the $\Gfour$ ExFT, described by the generalised metric $\mathcal{M}_{MN}$, which here can be written as $\mathcal{M}_{ab,cd} = 2 m_{a[c} m_{d]b}$ in terms of a symmetric unit determinant ``little metric'' $m_{ab}$ \cite{Duff:1990hn, Berman:2011cg}.
In the IIB solution of the section condition \cite{Blair:2013gqa}, we parametrise a generalised tensor $V^a$ as $V^a = ( V_i , V^\alpha)$, with $i=1,2,3$, and $\alpha=4,5$. This unusual convention for the $\mathrm{GL}(3)$ index $i$ is such that the extended coordinates are $Y^{ab} = ( Y_{ij} , Y_i{}^\alpha, Y^{\alpha \beta})$ with the physical coordinates $Y^i \equiv \frac{1}{2} \epsilon^{ijk} Y_{jk}$ carrying the usual upper index.
Then we parametrise the little metric $m_{ab}$ by
\be
\begin{split}
m_{ab} &=\begin{pmatrix}
\delta^i_k & \tilde v^{i \gamma} \\
0 & \delta_{\alpha}^{\gamma}
\end{pmatrix} 
\begin{pmatrix} 
g^{3/5} g^{kl} & 0 \\ 0 & g^{-2/5} \cH_{\alpha \beta} 
\end{pmatrix}
\begin{pmatrix} 
\delta_l^j & 0 \\
\tilde v^{j \delta} & \delta^{\delta}_{\beta}
\end{pmatrix} \\ &
= 
\begin{pmatrix}
g^{3/5} g^{ij} + g^{-2/5} \cH_{\gamma\delta} \tilde v^{i \gamma} \tilde v^{j \delta} & g^{-2/5} \cH_{\beta \gamma} \tilde v^{i \gamma} \\
g^{-2/5} \cH_{\alpha \gamma} \tilde v^{j \gamma} & g^{-2/5} \cH_{\alpha \beta} 
\end{pmatrix} \,.
\end{split}
\label{mIIB}
\ee
Here $g^{ij}$ is the inverse spacetime metric, $\tilde v^{i\alpha} \equiv \frac{1}{2} \epsilon^{ijk} \tilde C_{jk}{}^\alpha$ corresponds to the RR/NSNS 2-form doublet, with $C_{ij}{}^{\alpha}=(C_{ij}, B_{ij})$, and 
\be
\cH_{\alpha \beta} = e^{\Phi} \begin{pmatrix} 1 & C_0 \\ C_0 & e^{-2\Phi} + C_0^2 \end{pmatrix} 
\label{cHIIB}
\ee
contains the dilaton and RR 0-form.

U-duality transformations act such that 
\be
m_{ab} \rightarrow (U^t)_a{}^c m_{cd} U^d{}_b \,,\quad
V^a \rightarrow (U^{-1})^a{}_b V^b \,.
\ee
Geometric U-dualities include shifts of the two-forms, $\tilde v^{i\alpha} \rightarrow \tilde v^{i\alpha} + \Omega^{i \alpha}$, and $\mathrm{GL}(3)$ coordinate transformations generated by 
\be
(U_\Omega)^a{}_b = \begin{pmatrix} \delta_i^j & 0 \\ \Omega^{j\alpha} & \delta^\alpha_\beta \end{pmatrix} \,,\quad
(U_A)^a{}_b = \begin{pmatrix} ( \det A )^{3/5} (A^{-1})_i{}^j & 0 \\ 0 & ( \det A)^{-2/5} \delta^\alpha_\beta \end{pmatrix} \,.
\label{Ugeo}
\ee
We also have non-geometric U-dualities which shift ``bivectors'', generated by 
\be
(U_\omega)^a{}_b = \begin{pmatrix} \delta_i^j & \omega_{i\beta} \\ 0 & \delta^\alpha{}_\beta \end{pmatrix} \,,
\label{Uomega}
\ee
and $\mathrm{SL}(2)$ S-dualities
\be
(U_S)^a{}_b = \begin{pmatrix} \delta_i^j & 0 \\ 0 & S^\alpha{}_\beta \end{pmatrix} 
\,,\quad
S^\alpha{}_\beta = \begin{pmatrix} d &  b \\ c & a \end{pmatrix}
\,,\quad ad-bc=1\,,
\label{Us}
\ee
such that $\tau \rightarrow (a\tau+b)/(c\tau+d)$ for $\tau=C_0 + i e^{-\Phi}$.

\subsection{$\Gfour$ U-folds in IIB}

We want to consider configurations where the exceptional geometry is patched by $\Gfour$ transformations. 
The stereotypical situation is that our fields depend on some periodic coordinate $\theta$, and we have a monodromy $m_{ab}(\theta+2\pi) = (U^t)_a{}^c m_{cd}(\theta) U^d{}_b$.
For instance, we could tread the well-worn path of considering a three-torus with flux of the NSNS two-form (as inspired by \cite{Kachru:2002sk, Shelton:2005cf} and here essentially following the duality chains in \cite{Blair:2014zba}): 
\be
ds_E^2 = \eta_{\mu\nu} dx^\mu dx^\nu + \delta_{ij} dy^i dy^j \,,\quad B_{12} = H y^3 \,, \quad e^\Phi = 1 \,.
\label{boring} 
\ee
We have written the 10-dimensional Einstein frame metric $\hat g_{\hat\mu\hat\nu}$ in a $7+3$ split of the coordinates, $\hat\mu= (\mu,i)$, with ``external'' directions $\mu=0,\dots,6$ and ``internal'' directions $i=1,2,3$, which is appropriate for the $\Gfour$ ExFT. Our D-brane conditions will give us information about branes wrapping the internal directions. 

The metric components appearing in the IIB generalised metric are $g_{ij} \equiv \hat g_{ij}$. 
In the absence of off-diagonal components, the combination $g_{\mu\nu} \equiv (\det g_{ij})^{1/5} \hat g_{\mu\nu}$ is invariant under $\Gfour$ U-duality transformations.

The little metric for the background \eqref{boring} is:
\be
m_{ab} = \begin{pmatrix} 
1 & 0 & 0 & 0 & 0 \\ 
0 & 1 & 0 & 0 & 0 \\ 
0 & 0 & 1+(Hy^3)^2 & 0 & Hy^3 \\ 
0 & 0 & 0 & 1 &0 \\ 
0 & 0 & Hy^3 & 0 & 1 \end{pmatrix} \,. 
\ee
We will generate a new background by U-dualising with the transformation
\be
U^a{}_b = \begin{pmatrix} 1 & 0 & 0 & 0 & 0 \\ 0 & 1 & 0 & 0 & 0 \\ 0 & 0 & 0 & 0 & 1 \\ 0 & 0 & 0 & 1 & 0 \\ 0 & 0 & -1 & 0 & 0 \end{pmatrix} \,,
\label{Ubuscher35}
\ee
which in fact amounts to T-dualising in the $y^1, y^2$ directions. 
This leads to the Einstein frame configuration (the quantity in square brackets is the string frame metric):\footnote{In terms of the string winding coordinates of the original background \eqref{boring}, the coordinates here are $y^1 \equiv \tilde y_1$ and $y^2 \equiv \tilde y_2$.}
\be
\begin{split}
ds_E^2 &= e^{-\Phi/2} \left[ \eta_{\mu\nu} dx^\mu dx^\nu + (dy^3)^2 + \frac{1}{1+(Hy^3)^2} ( (dy^1)^2 + (dy^2)^2 )\right] \,,
\\
B_{12} & = - \frac{Hy^3}{1+(Hy^3)^2}\,,
\\
 e^\Phi &= (1+(Hy^3)^2)^{-1/2} \,.
\end{split} 
\label{boringU}
\ee
This is non-geometric for $y^3 \rightarrow y^3 + 2\pi$, and transforms as a T-fold under an $SO(2,2)$ duality transformation embedded in $\mathrm{SL}(5)$ as a $U_\omega$ of the form \eqref{Uomega} with
\be
\omega_{i\alpha} = 2 \pi H \begin{pmatrix} 0 & 0 \\ 0 & 0 \\ 0 & -1 \end{pmatrix} \,.
\ee
Acting with S-duality \eqref{Us} with $b=1$, $c=-1$, $a=d=0$, on the configuration \eqref{boringU} trivially generates a genuine U-fold, 
\be
\begin{split}
ds_E^2 &= e^{\Phi/2}\left[ \eta_{\mu\nu} dx^\mu dx^\nu + (dy^3)^2 + \frac{1}{1+(Hy^3)^2} ( (dy^1)^2 + (dy^2)^2  ) \right] \,,
\\ 
C_{12} & =  -\frac{Hy^3}{1+(Hy^3)^2}\,,
\\
 e^\Phi &= (1+(Hy^3)^2)^{1/2} \,,
\end{split} 
\label{boringUS}
\ee
with the U-fold monodromy again of the form \eqref{Uomega} with
\be
\omega_{i\alpha} = 2 \pi H \begin{pmatrix} 0 & 0 \\ 0 & 0 \\ -1 & 0 \end{pmatrix} \,.
\ee
Alternatively, we could S-dualise the original configuration \eqref{boring}, and then act with \eqref{Ubuscher35}, leading to a configuration with a flat background metric, vanishing two-forms, and
\be
\mathcal{H}_{\alpha \beta} = \begin{pmatrix} 1 & Hy^3 \\ Hy^3 & 1+(Hy^3)^2 \end{pmatrix} \Rightarrow C_0 =  Hy^3 \,,\quad e^\Phi = 1 \,.
\ee
This is an S-fold in that the monodromy lies in the $\mathrm{SL}(2)$ S-duality subgroup, and amounts to shifting $C_0 \rightarrow C_0 + 2\pi H$.\footnote{This is the monodromy of the D7 brane. Thinking of \eqref{boring} as a simplified configuration inspired by the NS5 brane, the duality chain here can be seen as S-dualising to the D5 and then T-dualising to the D7.} 
If we act with the fundamental S-duality again, we get
\be
\mathcal{H}_{\alpha \beta} = \begin{pmatrix} 1+(Hy^3)^2 & -Hy^3 \\  -Hy^3 & 1  \end{pmatrix} \Rightarrow C_0 = -\frac{Hy^3}{1+(Hy^3)^2} \,,\quad e^\Phi = \frac{1}{1+(Hy^3)^2} \,,
\label{nongeoSfold}
\ee
which is a ``non-geometric'' S-fold with $a=d=1$, $b=0$ and $c=-2\pi H$.

All the above configurations are meant to be illustrative examples of these standard monodromies. 
One can generate more realistic backgrounds by starting with the solution for the NS5 brane in place of the three-torus with H-flux given in \eqref{boring}, smearing twice in transverse directions and then dualising as above. Such chains of dualities are also discussed in \cite{Blair:2014zba}, and lead to non-geometric exotic branes \cite{deBoer:2012ma}.

\subsection{Monodromies, string charges and D-branes}

Now let's discuss how to make statements about strings and D-branes in U-folds using the previous discussion.
In $\Gfour$ exceptional geometry, we needed to combine the string charge $q^a$ with the $\mathbb{Z}_2$ transformation $Z^a{}_b$ in order to define D-branes.
We suppose that we can apply this definition to backgrounds with non-trivial $\Gfour$ monodromies. 
Under the monodromy transformation $U^a{}_b$, we have $q^a \rightarrow U^a{}_b q^b$ and $Z^a{}_b \rightarrow (U^{-1})^a{}_c Z^c{}_d U^d{}_b$.
We want to see how this affects the Dirichlet boundary conditions defined using the $\mathbb{Z}_2$ transformation.
For consistency, we require that the monodromy lead to the same boundary conditions.\footnote{Technically one need only actually require \cite{Lawrence:2006ma} that there is some integer $m$ such that $(U^{-m})^a{}_c Z^c{}_d (U^m)^d{}_b$ imposes the same boundary conditions. Our examples will only feature $m=1$.}

In the IIB case, the non-zero components of the string charge lay in the $\mathrm{SL}(2)$ directions, thus $q^a = ( 0 , q^\alpha)$. 
This is trivially preserved (up to a scaling in the case of $U_A$) under geometric U-dualities \eqref{Ugeo}.
Let us therefore consider the more interesting situations where we have monodromies lying in the S-duality subgroup or of the non-geometric type \eqref{Uomega}.

Under S-dualities, we have $q^\alpha \rightarrow S^\alpha{}_\beta q^\beta$.
The $\mathbb{Z}_2$ leading to D7 branes wrapping a single direction of the internal space is $Z^a{}_b = \text{diag}(-1,-1,+1,-1,-1)$.
This is clearly preserved by S-duality transformations \eqref{Us}.
Altogether the pair $(q^\alpha, Z^a{}_b)$ transform under an S-duality monodromy to $(S^\alpha{}_\beta q^\beta, Z^a{}_b)$. 
One can consider this as telling us that the 7-brane on which $(p,q)$ strings with charge $q^\alpha$ end is transformed into the 7-brane on which the $(p,q)$ strings with charge $S^\alpha{}_\beta q^\beta$ end.
This is what one would expect. 

The $\mathbb{Z}_2$ leading to 9-branes  
wrapping all three directions of the internal space is $Z^a{}_b = \text{diag}(-1,-1,-1,\pm 1, \mp 1)$.
The S-duality transformation \eqref{Us} turns this into
\be
Z^a{}_b = \begin{pmatrix} -1 & 0 & 0 & 0 & 0 \\ 0 & - 1 & 0 & 0 & 0 \\ 0 & 0 & -1 & 0 & 0 \\ 0 & 0 & 0 & \pm (1+ 2 bc ) & \pm 2 a b \\ 0 & 0 & 0 & \mp 2cd & \mp (1+ 2bc)\end{pmatrix} \,.
\ee
This leads to a Dirichlet projector such that (we label $a=(i,4,5)$ and do not lower the $i=1,2,3$ indices for convenience)
\be
\begin{split} 
(\Pd \delta Y)^{ij} & = 0 \,,\\
(\Pd \delta Y)^{i4} & = \frac{1}{2} ( \delta Y^{i4} \pm (1+2bc) \delta Y^{i4} \pm 2 ab \delta Y^{i5} ) \,,\\
(\Pd \delta Y)^{i5} & = \frac{1}{2} ( \delta Y^{i5} \mp (1+2bc) \delta Y^{i5} \mp 2 cd \delta Y^{i4} ) \,,\\
(\Pd \delta Y)^{45} &  = \delta Y^{45}\,.
\end{split}
\label{DirafterS}
\ee
With the upper sign, the original Dirichlet condition was $\delta Y^{i4} = 0 = \delta Y^{45}$.
The transformed Dirichlet conditions \eqref{DirafterS} are equivalent to these only if $ab=0=bc$.
Conversely, with the lower sign, the original Dirichlet condition was $\delta Y^{i5} = 0 = \delta Y^{45}$.
The transformed Dirichlet conditions are then equivalent only if $bc=0=cd$.
For instance, this tells us in the latter case (lower sign) that we cannot have branes wrapping the internal space in the non-geometric S-fold \eqref{nongeoSfold} for which $cd \neq 0$. However, they would be allowed in the former case (upper sign), as $b=0$. Additonally, the monodromy will lead to a mixed type of string/brane combination (that is, some mixture of F1/D9 and D1/NS9) unless $q^\alpha$ is preserved.

Next, we consider the non-geometric U-dualities \eqref{Uomega}.
Under these, we have $q^a \rightarrow ( \omega_{i\beta} q^\beta , q^\alpha )$.
The transformed charge $q^a$ will in this case not obey the charge condition $q^b \partial_{ab} = 0$ unless $\omega_{[i|\alpha} q^\alpha \partial_{j]} = 0$.
Dualising the indices on $\omega_{i\alpha}$, this is the same as $q^\alpha \omega^{ij}{}_\alpha \partial_k = 0$.
In this dualised form, $\omega^{ij}{}_\alpha$ can be seen as the shift in a bivector $\tilde C^{ij}{}_\alpha$. 
For generic $q^\alpha$, this tells us that in order to have well-defined strings we need the indices $i,j$ for which the bivector has non-zero components to correspond to isometry directions.
(This condition is frequently used for the bivector in the NSNS sector \cite{Andriot:2011uh}.) Observe that this is the case in the backgrounds \eqref{boringU} and \eqref{boringUS}, for which we have $\omega^{12}{}_\alpha \neq 0$, after T-dualising on the $y^1$ and $y^2$ directions, which were isometries.

We now consider the action of $U_\omega$ on the transformation $Z^a{}_b$. In general, we find that the monodromy turns $Z^a{}_b$ into
\be
Z^{\prime a}{}_b \equiv ( U_\omega^{-1} Z U_\omega )^a{}_b = \begin{pmatrix} Z_i{}^j & Z_i{}^k \omega_{k \beta} - \omega_{i \gamma} Z^\gamma{}_\beta \\ 0 & Z^\alpha{}_\beta \end{pmatrix} \,.
\label{transformedZ}
\ee
Clearly, $Z^a{}_b$ will be preserved if the top-right block vanishes, so that the boundary conditions are trivially invariant under the monodromy.
For the D7 case, we have $Z^\alpha{}_\beta = -\delta^\alpha{}_\beta$, and $Z^i{}_j$ acting as $-1$ in two directions and $+1$ in the other direction.
Then the top-right block of \eqref{transformedZ} is non-zero if $\omega_{i_+ \alpha} \neq 0$, where $i_+$ denotes this even direction.
For the D9 case, we have $Z^i{}_j = -\delta^i{}_j$ and $Z^\alpha{}_\beta = \text{diag}(\pm 1 , \mp 1 )$, so this is non-vanishing if either $\omega_{i 4}$ or $\omega_{i 5}$ is non-vanishing, depending on the sign choice.

Let us look at the D7 case in more detail. We consider the case $Z^i{}_j = \text{diag}(-1,-1,+1)$ corresponding to a brane wrapping the $Y^{12}$ direction, and the monodromy determined by $\omega_{3\alpha} \equiv (\omega, \tilde \omega)$, which can describe the examples \eqref{boringU} and \eqref{boringUS} (in which $Y^{12} \equiv y^3$). The transformed $Z^a{}_b$ is 
\be
Z^{\prime a}{}_b = \begin{pmatrix} 
-1 & 0 & 0 & 0 & 0 \\ 
0 & - 1 & 0 & 0 & 0 \\ 
0 & 0 & +1 & 2\omega & 2\tilde \omega \\ 
0 & 0 & 0 & -1 & 0  \\ 
0 & 0 & 0 & 0 & -1 \\ 
\end{pmatrix} \,.
\ee
We again analyse the Dirichlet projection condition, finding that
\be
(\Pd \delta Y)^{12}  = (\Pd \delta Y)^{14}=(\Pd \delta Y)^{24}=(\Pd \delta Y)^{15}=(\Pd \delta Y)^{25}=(\Pd \delta Y)^{45} = 0\,,
\ee
and
\be
\begin{split} 
(\Pd \delta Y)^{13} & = \delta Y^{13} + \omega \delta Y^{14} + \tilde \omega \delta Y^{15} \,,\\
(\Pd \delta Y)^{23} & = \delta Y^{23} + \omega \delta Y^{24} + \tilde \omega \delta Y^{25} \,,\\
(\Pd \delta Y)^{34} &  = \delta Y^{34}+ \tilde \omega \delta Y^{45} \,,\\
(\Pd \delta Y)^{35} &  = \delta Y^{35}-  \omega \delta Y^{45} \,.
\end{split}
\label{DirafterS2}
\ee
The original Dirichlet projection sets $\delta Y^{13} = \delta Y^{23} = \delta Y^{34} = \delta Y^{35} = 0$. Thus this describes a brane wrapping the direction $Y^{12}$ of the internal space, and fixed in the directions $Y^{13}$ and $Y^{23}$. 
The transformed projection \eqref{DirafterS2} is however inequivalent. 
This rules out D1 branes wrapping the direction $y^3$ in the T-fold background \eqref{boringU}, which is the ``base'' direction, in agreement with \cite{Lawrence:2006ma}, and also in the S-dual.
Note though that for the same monodromy, the cases $Z^i{}_j = \text{diag}(+1,-1,-1)$ and $Z^i{}_j = \text{diag}(-1,+1,-1)$, corresponding to branes wrapping just the $y^1$ and $y^2$ directions, are allowed ($Z^a{}_b$ is invariant), as found in \cite{Albertsson:2008gq}.

Now examine the D9 case, for 
\be
Z^a{}_b = \begin{pmatrix} 
-1 & 0 & 0 & 0 & 0 \\ 
0 & - 1 & 0 & 0 & 0 \\ 
0 & 0 & -1 & 0 & 2\tilde \omega \\ 
0 & 0 & 0 & -1 & 0  \\ 
0 & 0 & 0 & 0 &  + 1 \\ 
\end{pmatrix} \,,
\ee
where we take $\omega_{i \alpha}$ to have only one non-vanishing component in order to preserve the string charge (assuming isometries as above).
We yet again analyse the Dirichlet projection condition, finding that
\be
(\Pd \delta Y)^{12}  = (\Pd \delta Y)^{14}=(\Pd \delta Y)^{24} = 0\,,
\ee
and
\be
\begin{split} 
(\Pd \delta Y)^{13} & = \tilde \omega \delta Y^{15} \,,\\
(\Pd \delta Y)^{23} & =  \tilde \omega \delta Y^{25} \,,\\
(\Pd \delta Y)^{45} & =   \delta Y^{45} \,,\\
(\Pd \delta Y)^{15} & =  \delta Y^{15} \,,\\
(\Pd \delta Y)^{25} & =   \delta Y^{25} \,,\\
(\Pd \delta Y)^{34} &  =-\tilde \omega \delta Y^{45} \,,\\
(\Pd \delta Y)^{35} &  = \delta Y^{35} \,.
\end{split}
\label{DirafterS3}
\ee
These are consistent with the original projection conditions: therefore we can have three-branes in the T-fold background \eqref{boringU}, again agreeing with \cite{Lawrence:2006ma}. 
S-duality interchanges the $4$ and $5$ indices, and shows that these three-branes are also possible in the U-fold S-dual to \eqref{boringU}. 

It is clear how to continue this analysis for other monodromies, and also in the IIA case.
We hope that the above discussion demonstrates the general situation adequately. 

\section{Discussion}

In this short paper, we have scratched the surface of the topic of D-branes, and some of their S-duals, in exceptional geometry. This involved combining previous work on strings whose target space is this exceptional geometry \cite{Arvanitakis:2017hwb, Arvanitakis:2018hfn} with the study of generalised orientifolds in ExFT \cite{Blair:2018lbh}, providing a promising route in to the study of D-branes in this setting. 
We would like to propose a number of developments one could now attempt building on this work.

\vspace{1em}
\emph{More on D-branes.} We only studied the simplest examples of D-branes in this paper. One could say much more about their presence or absence in U-folds. For instance, we did not consider locally non-geometric examples, where the background spacetime depends explicitly on dual coordinates.
The description of D-branes in T-folds was recently revisited in \cite{Hull:2019iuy, Lust:2019hmr} in order to take decoupling limits leading to non-commutative and non-associative theories on the D-brane worldvolume: we should explore how the obvious generalisations to U-folds may work.

We briefly mentioned the possibility of having different boundary conditions at the string endpoints, attaching each end of the string to separate subspaces of the exceptional geometry. This would involve a pair of projectors, $\Pd, \widetilde{\Pd}$, each compatible with the same string charge $q$ but in general preserving different half-maximal structures. Overall this would generically give a configuration with less SUSY. There is then likely a neat classification of such intersecting brane configurations available with this approach.

It might also be possible to study D-branes in so-called non-Riemannian backgrounds \cite{Morand:2017fnv, Berman:2019izh}, where the generalised metric cannot be parametrised in terms of an invertible spacetime metric: this might allow a novel way to define D-branes in non-relativistic theories, for instance.

\vspace{1em}
\emph{Heterotic strings?} In the IIB case, our definition led also to the S-duals of the usual D-branes. This included not only the $(p,q)$ 7-branes, but also an S-dual of the D9 brane. This would be an NS9 brane on which open D1 branes end. This should correspond in fact to the heterotic $\mathrm{SO}(32)$ string, and so it is natural to ask whether the open string version of the exceptional sigma model provides a novel and perhaps unexpected duality symmetric treatment of the type I and heterotic strings, combining insights from this paper with the results of \cite{Hull:1998he, Blair:2018lbh}.

\vspace{1em}
\emph{Branes ending on branes.} We had string charges $q$ obeying the constraint $q \otimes \partial |_{R_3} = 0$, which we solved assuming the derivative $\partial_M$ corresponded to the solutions of the section condition giving 10-dimensional IIA or IIB. 
In principle, if we assume isometries, so that $\partial_M = 0$, then the charge $q$ is unconstrained, and describes the full $E_{d(d)}$ multiplet of strings obtained in $(11-d)$ dimensions by partially wrapping branes on a $T^{d-1}$ torus. In this case, the definition for D-branes we used may also describe the embedding into the exceptional geometry of the more general set of branes on which these partially wrapped branes can end. 

One could potentially also proceed to study higher rank branes directly. For instance, membranes in exceptional geometry must be characterised by a charge $\tilde q \in \bar{R}_3$, obeying constraints such as $\tilde q \otimes \partial |_{R_4} = 0$. Requiring for example $\tilde q_{MNP} = - Z^{K}{}_M Z^L{}_N Z^{Q}{}_P \tilde q_{KLQ}$ with the same $\mathbb{Z}_2$ may then allow us to obtain exceptional geometric definitions of the branes on which membranes end. Indeed, for $\mathrm{SL}(5)$ this charge can be seen to be $\tilde q_a$, obeying $\tilde q_{[a} \partial_{bc]} =0$. In the M-theory solution of the section condition we have $\partial_{i5} \neq 0$, $i=1,2,3,4$, so that $\tilde q_5 \neq 0$, $\tilde q_i = 0$, informing us of the existence of the M2. Requiring $Z^a{}_b \tilde q_a = - q_b$, this is compatible with a $\mathbb{Z}_2$ transformation of the form $Z^a{}_b = \text{diag}( -1,-1,-1,+1,-1)$, such that the projectors imply the brane wraps the three directions $Y^{15}, Y^{25}, Y^{35}$ in the internal space. This suggests it is an M9 brane, as in Ho\v{r}ava-Witten \cite{Horava:1995qa, Horava:1996ma}, on which M2 branes do end.

\vspace{1em}
\emph{Brane actions.} It would be interesting to formulate fully $E_{d(d)}$ covariant actions for the branes discussed in this paper. 
Some approaches to D-branes in the $O(D,D)$ case which may be applicable include \cite{Albertsson:2011ux, Asakawa:2012px, Bergshoeff:2019sfy} (see also \cite{Blair:2017hhy} for (NS)5-branes in DFT and \cite{Sakatani:2017vbd} in ExFT).
In fact, the paper \cite{Bergshoeff:2010xc} has already described U-duality covariant expressions for the Wess-Zumino terms of D-branes in various dimensions. 
Interestingly, this involved a doubling of the number of worldvolume scalars corresponding to internal directions. 
In an approach based on the exceptional geometry of ExFT, we would want to embed these doubled coordinates into the full extended coordinates $Y^M$ (this may be reminiscent of how the exceptional sigma model contains a reduction to the doubled sigma model \cite{Arvanitakis:2018hfn}), and also to understand the $Y^M$-dependent gauge transformations of the generalised gauge field to which the brane will couple electrically.
The most natural case to consider is that of branes which are external spacetime filling, and so couple to an $E_{d(d)}$ multiplet of forms $\mathcal{C}_{\mu_1\dots \mu_n}$ which lies beyond the usual tensor hierarchy construction needed in ExFT. 
The $E_{d(d)}$ representations of these forms and the structure of the charges to which they couple (the generalisations of the string charge $q$ appearing in the exceptional sigma model) have been described in \cite{Bergshoeff:2010xc}.

\vspace{1em}
\emph{More on the geometry.}
We would also like to obtain a more comprehensive understanding of the geometry of the subspaces defined by our projectors. 
The numerology of the number of Dirichlet directions is quite appealing (see table \ref{nos}) in this regard. 
There may also be more to say about the interplay between the string charge $q$ and the half-maximal structure (see appendix \ref{extra}).
Another observation is the following. 
In doubled geometry, one can view the $D$-dimensional D-brane as well as the physical subspace as maximally isotropic subspaces of the $2D$-dimensional space. This way of viewing the physical subspace is important for generalised dualities using the notion of a Drinfeld double \cite{Klimcik:1995ux}. 
Perhaps similar structures, and generalised generalised dualities, are implied by the branes in exceptional geometry.

\section*{Acknowledgements} 

I am supported by an FWO-Vlaanderen Postdoctoral Fellowship, and in part by the FWO-Vlaanderen through the project G006119N and by the Vrije Universiteit Brussel through the Strategic Research Program ``High-Energy Physics''. I would like to thank Emanuel Malek and Daniel Thompson for helpful conversations.

\appendix 

\section{$\Gfour$ IIB $(p,q)$ strings} 

This is a check on our conventions for labelling the $\mathrm{SL}(2)$ indices on the IIB solution of the section condition in $\Gfour$.
The tension \eqref{tension} of the $\Gfour$ exceptional sigma model is $T = \sqrt{m_{ab} q^a q^b}$.
In the IIB parametrisation given by \eqref{mIIB} and \eqref{cHIIB}, this becomes, with $q^a=(0, q^{\dalpha})$, $\dalpha=4,5$,
\be
T = \sqrt{ ( q^4 + C_0 q^5 )^2 + e^{-2\Phi} (q^5)^2} e^{\Phi/2} (\det g_{ij})^{-1/5}\,.
\ee
After integrating out the non-zero components of the auxiliary worldsheet field $\gV_A^M$, the exceptional sigma model becomes ($\hat \mu, \hat \nu$ are 10-dimensional indices, $\hat g_{\hmu\hnu}$ is the Einstein frame metric and $\hat B_{\hmu\hnu \dalpha}$ is a doublet of two-forms in 10 dimensions):
\be
\begin{split} 
S = - \frac{1}{2} \int d^2\sigma& \sqrt{-\gamma} \gamma^{A B} \sqrt{ ( q^4 + C_0 q^5 )^2 + e^{-2\Phi}(q^5)^2 } e^{\Phi/2} \hat g_{\hmu \hnu} \partial_A X^{\hmu} \partial_B X^{\hnu} 
\\ &+\epsilon^{AB} q^{\dalpha} \hat B_{\hmu\hnu \dalpha} \partial_A X^{\hmu} \partial_B X^{\hnu}\,.
\end{split}
\ee
We see that $q^4 \neq 0, q^5 = 0$ gives the fundamental string, while $q^5 \neq 0 , q^4 = 0$ gives a D1 action. 
This requires $\hat B_{\hmu\hnu \dalpha} = \epsilon_{\dalpha\dbeta} \hat B_{\hmu\hnu}{}^{\dbeta}$ with $\hat B_{\hmu \hnu}{}^{\dalpha} = (\hat C_{\hmu\hnu} , \hat B_{\hmu \hnu} )$.

\section{Comments on half-maximal structures and $O(D,D)$}
\label{extra} 

Our definition of the D-brane structure in exceptional geometry in section \ref{defD} included compatibility with a Neumann projected half-maximal structure, involving $J_u \in \Gamma(\mathcal{R}_1)$ and $K \in \Gamma(\mathcal{R}_2)$, such that $q(J_u, J_v ) = 0$, $q\cdot K = 0$.
In \cite{Malek:2017njj}, $O(D,D)$ half-maximal structures are discussed. These correspond to particular embeddings of $O(D) \subset O(D) \times O(D) \subset O(D,D)$.
The paper \cite{Malek:2017njj} looked in detail at a half-maximal structure $J_u^M$ such that $\eta_{MN} J_u^M J_v^N = \delta_{uv}$.
As $q_{MN} = \eta_{MN}$ in this case, this is naively at odds with our definition.
However, the choice of the $J_u^M$ corresponds to identifying $O(D)$ with one of the factors in the denominator subgroup $O(D)\times O(D)$, and allows them to be interpreted as the left-moving generalised vielbein. There are also the right-moving vielbein $\tilde J_u^M$ obeying $\eta_{MN} \tilde J_u^M \tilde J_v^N =- \delta_{uv}$. 
We conjecture that what seems natural for the D-brane structure, based on the form of supercharges preserved by open string boundary conditions on the worldsheet, would be to consider linear combinations $J_u^M + \tilde J_u^M$ of putative left- and right-moving half-maximal structures, i.e. a diagonal embedding into $O(D) \times O(D)$.
In the absence of a B-field, one can take these vielbein to be
\be
J_u^M = \frac{1}{\sqrt{2}}\begin{pmatrix} e_u{}^i \\ e_{ui} \end{pmatrix} 
\,\quad
\tilde J_u^M = \frac{1}{\sqrt{2}}\begin{pmatrix} \tilde e_u{}^i \\ -\tilde e_{ui} \end{pmatrix} \,,
\ee
in terms of separate left and right vielbein for the spacetime metric. Fixing $e=\tilde e$, the sum and difference
\be
J_u^M + \tilde J_u^M = \sqrt{2} \begin{pmatrix} e_u{}^i \\ 0 \end{pmatrix} \,,\quad
J_u^M - \tilde J_u^M = \sqrt{2} \begin{pmatrix} 0 \\ e_{ui} \end{pmatrix} 
\ee
are naturally Neumann and Dirichlet projected for $\Pn = \text{diag}(I,0)$, $\Pd = \text{diag}(0,I)$, corresponding to a spacetime filling D9 brane.
Indeed, in flat backgrounds at least, one can consider then combinations $J_u^M + \mathcal{P}^M{}_N \tilde J_u^N$ where $\mathcal{P}^M{}_N \in O(D,D)$ is the geometric transformation acting as a reflection in $D-p-1$ directions, in order to describe $p$-branes. 

It is also interesting to consider the explicit reduction to $O(D,D)$ of our definition. 
Consider again $\Gfour$, let the 5-dimensional fundamental index $a=(i,4,5)$ and fix $q^4 \neq 0$, $q^i = q^5 = 0$, so that the string charge defines F1 strings (in both IIA and IIB). 
Because we require $Z^a{}_b q^b = - q^a$ but $Z^b{}_a K_b = K_b$, this means we have $K_4 = 0$. 
Normally, the idea is to fix $K_{\sharp} \neq 0$, $K_\alpha = 0$, for $a=(\alpha,\sharp)$, such that the compatibility condition $J_u \wedge J_v = \delta_{uv} K$ becomes
\be
\delta_{uv} K_\sharp = \frac{1}{4} \epsilon_{\sharp \alpha \beta \gamma \delta} J_u^{\alpha \beta} J_v^{\gamma \delta} \,,\quad
0 = J_{(u}^{[\alpha \beta} J_{v)}^{\gamma] \sharp}\,.
\ee
Then picking $J_u^{\alpha \sharp} = 0$, and rescaling $J_u^{\alpha \beta}$ by a power of $K_{\sharp}$ (which is proportional to the generalised dilaton $e^{-2d}$), this is equivalent to $J_u^M J_v^N \eta_{MN} = \delta_{uv}$ after splitting $\alpha = ( i , 5)$ with $J_u^M = ( J_u^{i5}, J_u^{ij})$, $i,j=1,2,3$.
This gives the $O(D,D)$ half-maximal structure selected in \cite{Malek:2017njj}.
For us, the ``physical'' coordinates are fixed by the choice of $q^a$ (via the string charge constraint \eqref{magic}) and the $O(D,D)$ structure determined by the choice of which component $K_{\sharp}$ is non-zero will not coincide with the natural $E_{d(d)} \rightarrow (S)O(D,D)$ picked out by the coordinates. 

For instance, if $K_5 \neq 0$, $K_i = K_4 = 0$, we find the $\Gfour$ generalised vector $J_u^{ab}$ reduces to a non-zero $O(D,D)$ vector and spinor, with 
\be
J_u^M = \begin{pmatrix} 0 \\ J_u^{ij} \end{pmatrix}
\,,\quad
J_u^I = \begin{pmatrix} 0 \\ J_u^{i4} \end{pmatrix}
\label{JJ}
\ee
($I$ is a four component spinor index). The compatibility constraints are now
\be
J_u^M J_v^M \eta_{MN} = \delta_{uv} K_5 \,,\quad \gamma_{M IJ} J_{(u}^M J_{v)}^J \sim \delta_{uv} K_I  \,,
\label{JJKODD}
\ee
involving an off-diagonal block of the $O(3,3)$ gamma matrices.
As $J_u^M$ in \eqref{JJ} is meant to be Neumann projected, we can see that when the physical coordinates are $Y^{ij}$, i.e. in IIB, this corresponds to a D9 brane (because $\Pn = \text{diag}(I,0)$), while when the physical coordinates are $Y^{i5}$, i.e. in IIA, this corresponds to a D6 brane (because $\Pn = \text{diag}(0,I)$).
Taking $K_i \neq 0$ gives again the conditions \eqref{JJKODD} with particular forms of $J_u$ and $K$ corresponding to the D7 and D8 cases.

\bibliography{NewBib}

\end{document}